\documentclass{moriond}

\def\Journal#1#2#3#4{{#1} {\bf #2}, #3 (#4)}

\def\JHEP{{\em Journal of High Energy Physics}}

\def\ra{\rightarrow}

\def\bb{\mbox{$b\overline{b}$}}
\def\tt{\mbox{$t\overline{t}$}}
\def\hbb{\mbox{$H \ra \bb$}}
\def\ttH{\mbox{\tt$H$}}
\def\ifb{\mbox{fb$^{-1}$}}
\def\etmiss{E$_{\mathrm T}^{\mathrm miss}$}
\def\pT{p$_{\mathrm T}$}

\def\be{\begin{equation}}
\def\ee{\end{equation}}
\def\bea{\begin{eqnarray}}
\def\eea{\end{eqnarray}}

\newcommand{\Photo}{\includegraphics[height=35mm]{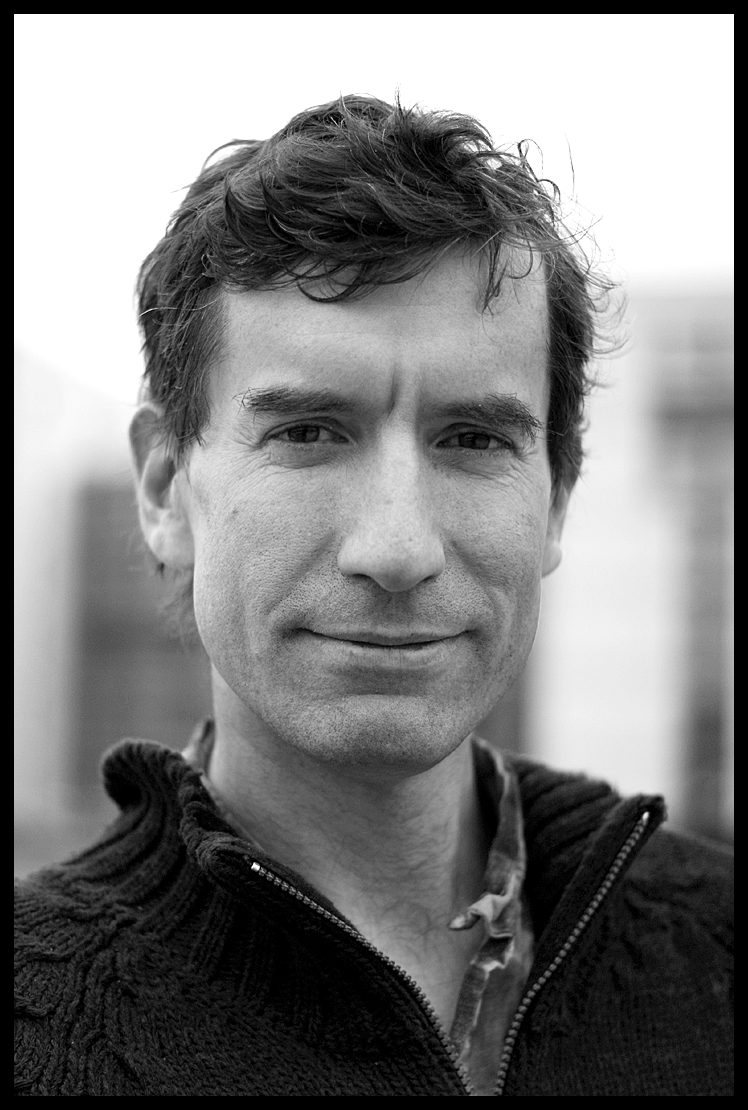}}

\begin{document}
\vspace*{4cm}
\title{Higgs with Hadronic Signatures}

\author{ W.J. Murray, on Behalf of the ATLAS and CMS collaborations }

\address{Department of Physics, Warwick University, CV4 7AL, UK \\ \&
  Rutherford Appleton Lab., Didcot, OX11 0QX, UK}

\maketitle\abstracts{
The decay of the Higgs boson to b quarks should be the dominant decay
mode, but it has not yet been conclusively established. The LHC run 1
results are recalled and the current
status of the LHC Run 2  studies is reviewed. The  analysis is
approaching decisive sensitivity.}

\section{Introduction}

The  ATLAS and CMS combined  Higgs coupling results\cite{ac2012} from
LHC Run 1 included \hbb\ searches in association with vector bosons and top
quarks. When Standard Model
production is assumed then  Br(\hbb) is
$\mu=0.70^{+0.29}_{-0.27}$, where $\mu$ is measured over
expected. If both production and decay
rates are fitted,  then Br(\hbb)/Br($H\ra ZZ$) has $\mu=0.20^{+0.20}_{-0.12}$.

This report discusses the  results   on \hbb\ from Run 2,
which are currently a fragmentary picture but give indications of the
direction. Cross-sections  rise as the beam energy goes from 8~TeV to
13~TeV, especially in \ttH\  where the (expected) number of Higgs
events analysed exceeds Run 1. The analyses have large
backgrounds, so machine learning techniques such as Boosted
Decision Trees (BDTs)\cite{bdt} are  used  to identify
events  with enhanced signal to background.

VBF production is studied by
CMS\cite{cms-vbf} and ATLAS\cite{atlas-vbf}, and is discussed in
section~\ref{sec:vbf}. VH production from ATLAS is in
section~\ref{sec:vh} and \ttH\ from both
experiments\cite{cms-tth,atlas-tth} is in
section~\ref{sec:tth}. CMS tH~\cite{cms-th} is  omitted for space
reasons. Finally section~\ref{sec:conclusions} contains  some conclusions.

\section{VBF Higgs production}\label{sec:vbf}

The highest-rate Higgs production process used for \hbb\ is vector
boson fusion, with a cross-section of 3.8~pb.
ATLAS and CMS have released results using   different approaches. CMS,
following Run 1, and with 2.3~\ifb,   used  a dedicated trigger selecting 
the topology of a central pair of b-jets and two forward jets, known
as tag jets. ATLAS instead, from 12.6~\ifb, use the much rarer
topology where  an 
additional photon is produced. This  simplifies the trigger and raises
the signal to background but with a large price in rate.

The CMS trigger requires 3 jets at L1, and four at the HLT. Two
categories are selected, either requiring a jet 
identified as a b-jet and a mass over 460~GeV for the tag pair, or two
b jets and only 200 GeV mass in the tag pair. The signal efficiency is
6\%. ATLAS used the photon as an L1 trigger, and then require a 700 GeV
tag jet pair mass at HLT with no b-tag requirement.

Both ATLAS and CMS rely on BDTs to enrich in signal purity. They use
different variables, but are sensitive to the same six areas: the
dynamics of the VBF system, the b-jet content of the event, the quark
or gluon nature of the tag jets, the suppression of jet activity
between the tag jets, the H production angles and the overall momentum
of the system of four jets. The modelling in the ATLAS case is
corrected using side-bands. The resulting BDT output is used to
define several signal regions.

\begin{figure}[ht]
  \centerline{
    \includegraphics[height=0.45\linewidth]{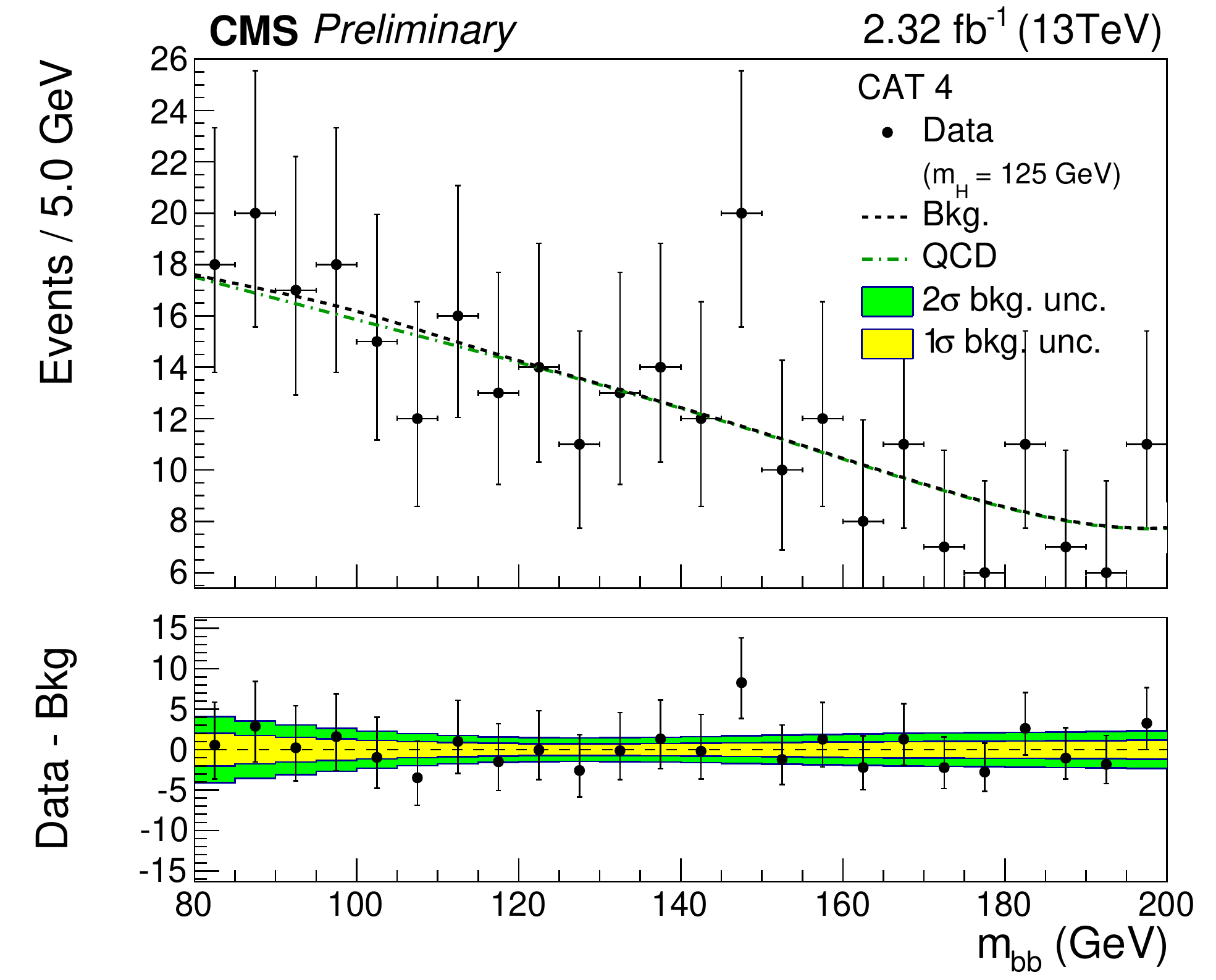}
    \includegraphics[height=0.45\linewidth]{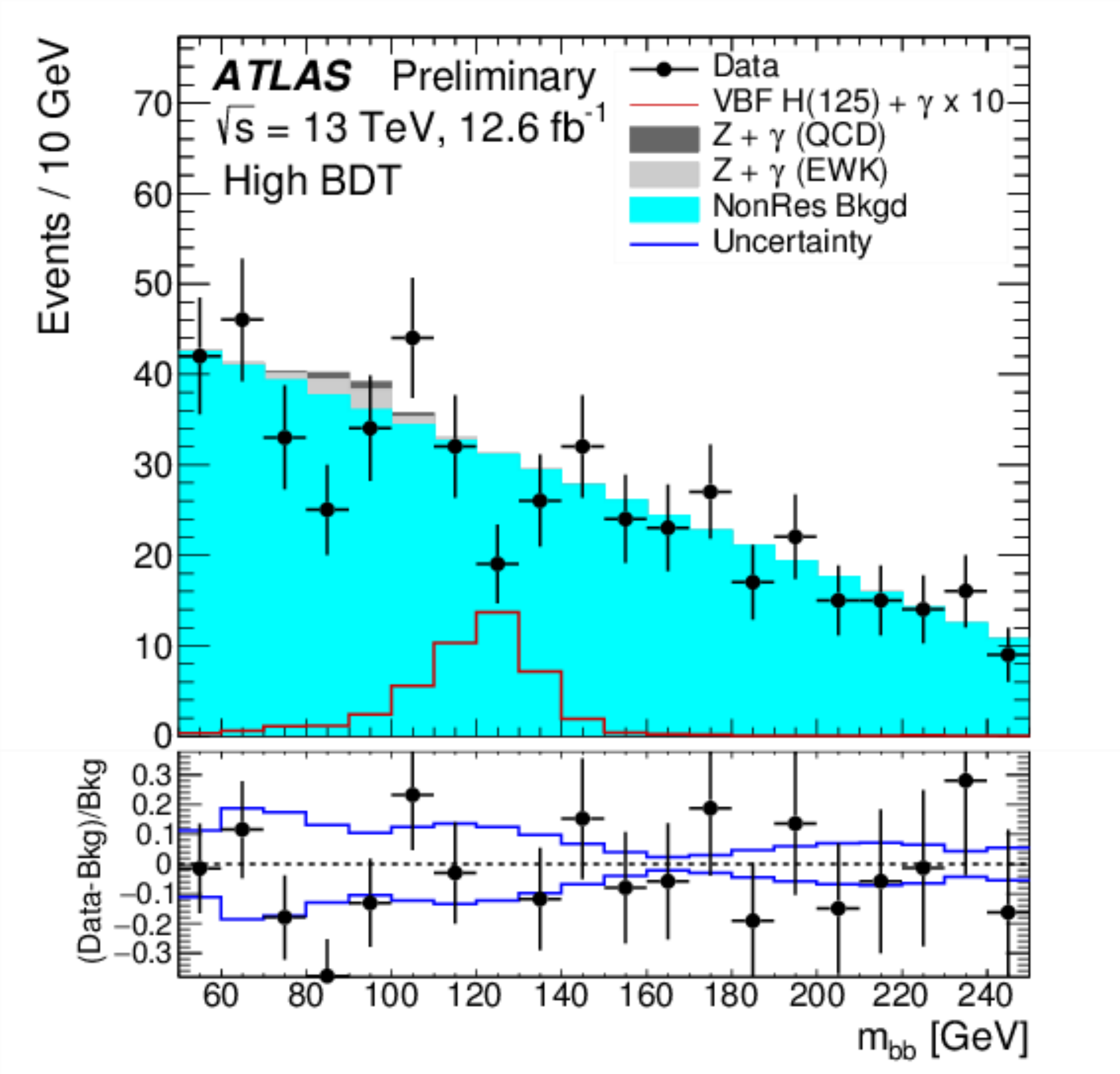}
  }
\caption[]{The most sensitive regions of the CMS (left) and ATLAS
  (right) VBF \hbb\ searches.}
\label{fig:vbf}
\end{figure}

The BDT selections are designed not to bias the mass of the Higgs dijet,
allowing the signal 
rate to be extracted from a fit to a smooth background and a signal
peak, which reduces dependence upon the background
simulation. In the CMS case the   background shape is constrained using the 
lower purity regions.  Figure~\ref{fig:vbf} shows the most sensitive regions. The resulting estimates of the signal strength
are $\mu=-3.7^{+2.4}_{-2.5}$ for CMS and $\mu=-3.9^{+2.8}_{-2.7}$ for ATLAS.

\section{Vector-boson associated }\label{sec:vh}

The total cross-sections for ZH and WH are 0.88~pb and 1.37~pb respectively,
but only non-hadronic vector boson decays are used. Furthermore, the
$V$\bb\ background in an inclusive analysis is very large, but  a
boosted topology provides the sensitivity. ATLAS have released
results\cite{atlas-vh} using 13.2~\ifb\ which are briefly reported here.

The analysis uses selections for  0, 1 or 2 leptons, (muons or electrons),
targeting $Z\ra \nu\overline{\nu}$, $W\ra l \nu$ and $Z\ra ll$
respectively. In each case two b-jets are required from the Higgs,
with possible extra jets.  The backgrounds and triggers vary considerably: two
leptons provides the cleanest, but low rate, channel; one lepton  has
the highest rate but large top backgrounds and the zero lepton
channel has the best overall sensitivity.

The trigger comes from either the charged leptons or \etmiss, and is
most problematic in the zero lepton channel, where an online threshold
of 90~GeV was used in 2016. All channels apply a \pT\ threshold of
150~GeV to the Higgs candidate, with the exception of the di-lepton,
where an additional signal region of lower \pT\ events is used.
A BDT is used to produce signal-enriched regions, but the mass is
included, implying that the background modelling is critical. Figure~\ref{fig:vh} shows
the impact on $\mu$ of various systematic errors; dominated by flavour
tagging and the normalisation of $Z$ plus heavy flavour background.

\begin{figure}[ht]
  \centerline{
    \includegraphics[height=0.4\linewidth]{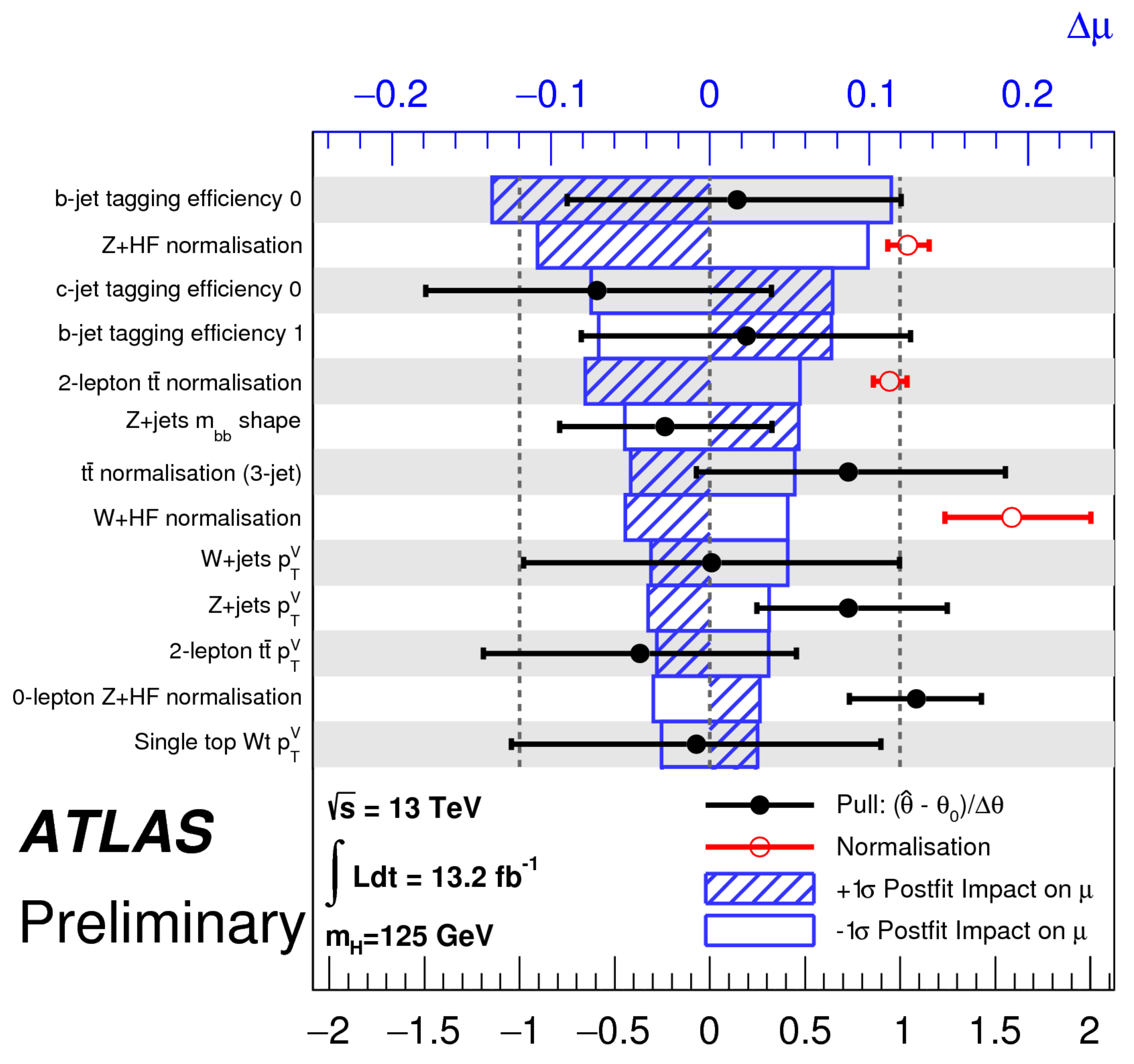}
    \includegraphics[height=0.4\linewidth]{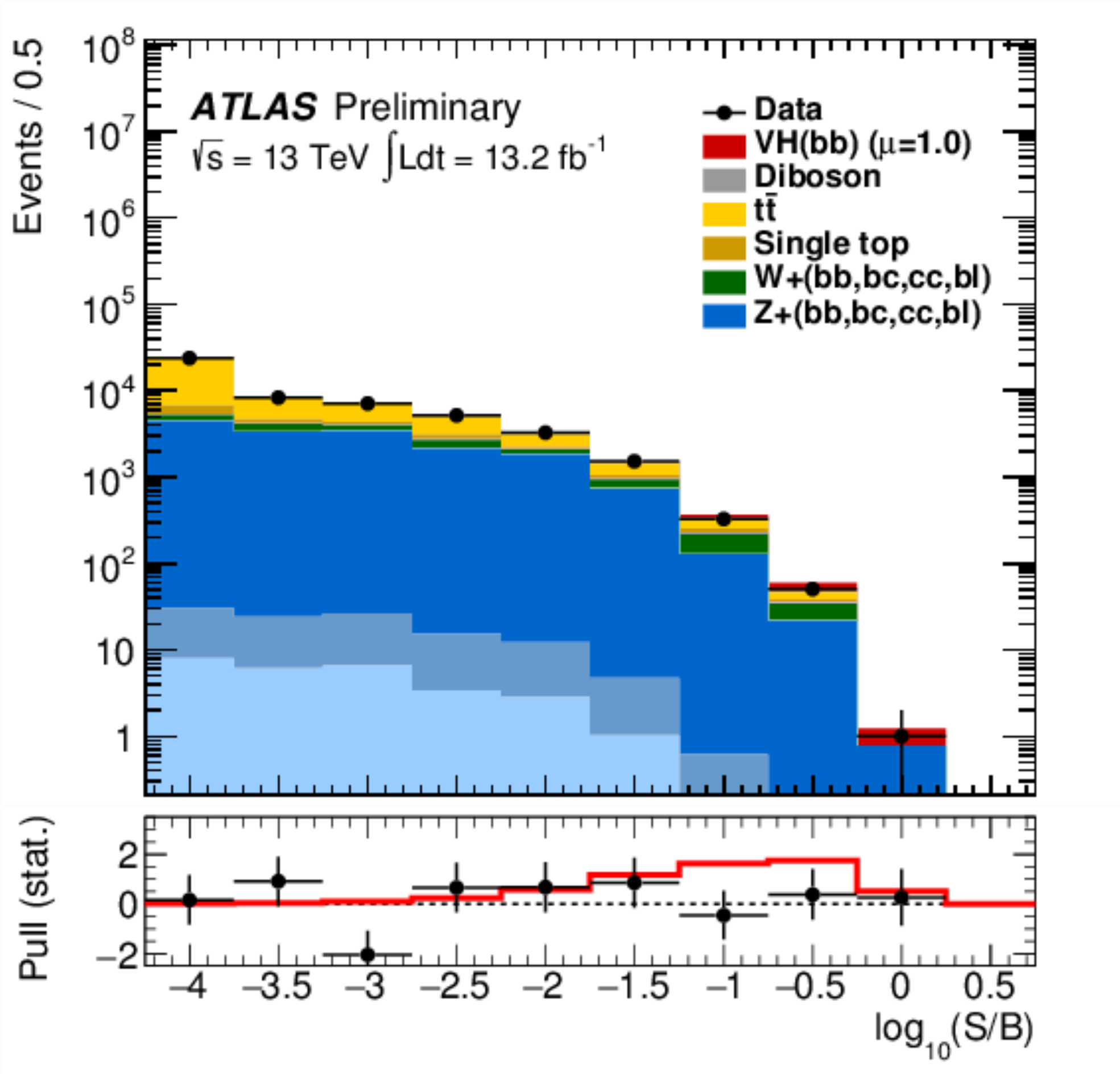}
  }
\caption[]{The VH analysis. Left is the leading constraints on nuisance
  parameters of the fit, and right  the combined BDTs of all the 
  VH channels.} 
\label{fig:vh}
\end{figure}

The resulting signal strength is $\mu= 0.21\pm 0.36(stat)\pm
0.36(sys)$. A test using  the same methodology, but
searching for $VZ$, yields $\mu=0.91\pm0.17(stat)\pm 0.30(sys)$,
giving confidence in the modelling.

\section{Top-pair associated }\label{sec:tth}

Both CMS and ATLAS have released
\ttH\ analyses\cite{cms-tth,atlas-tth} based on about 13~\ifb\ of data.
The \ttH\ channel  cross-section rose by a factor 3.8 from 8 TeV to
0.51~pb. The leptonic top decay provides an  excellent tag and trigger, but the
complex final state means that \tt, and especially \tt\bb, modelling
issues dominate the analysis. 

\begin{figure}[ht]
  \centerline{
    \includegraphics[width=0.45\linewidth]{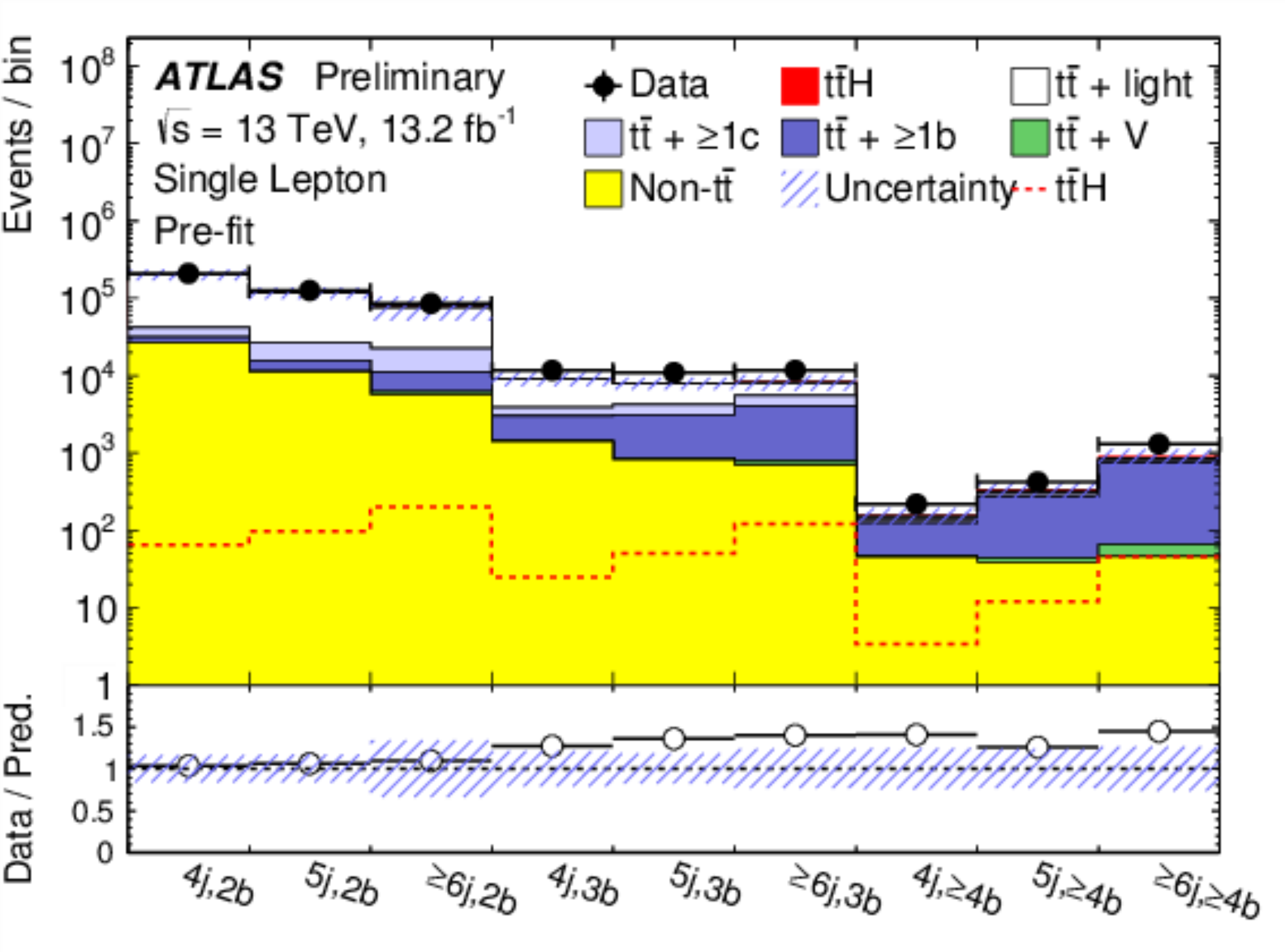}
    \includegraphics[width=0.45\linewidth]{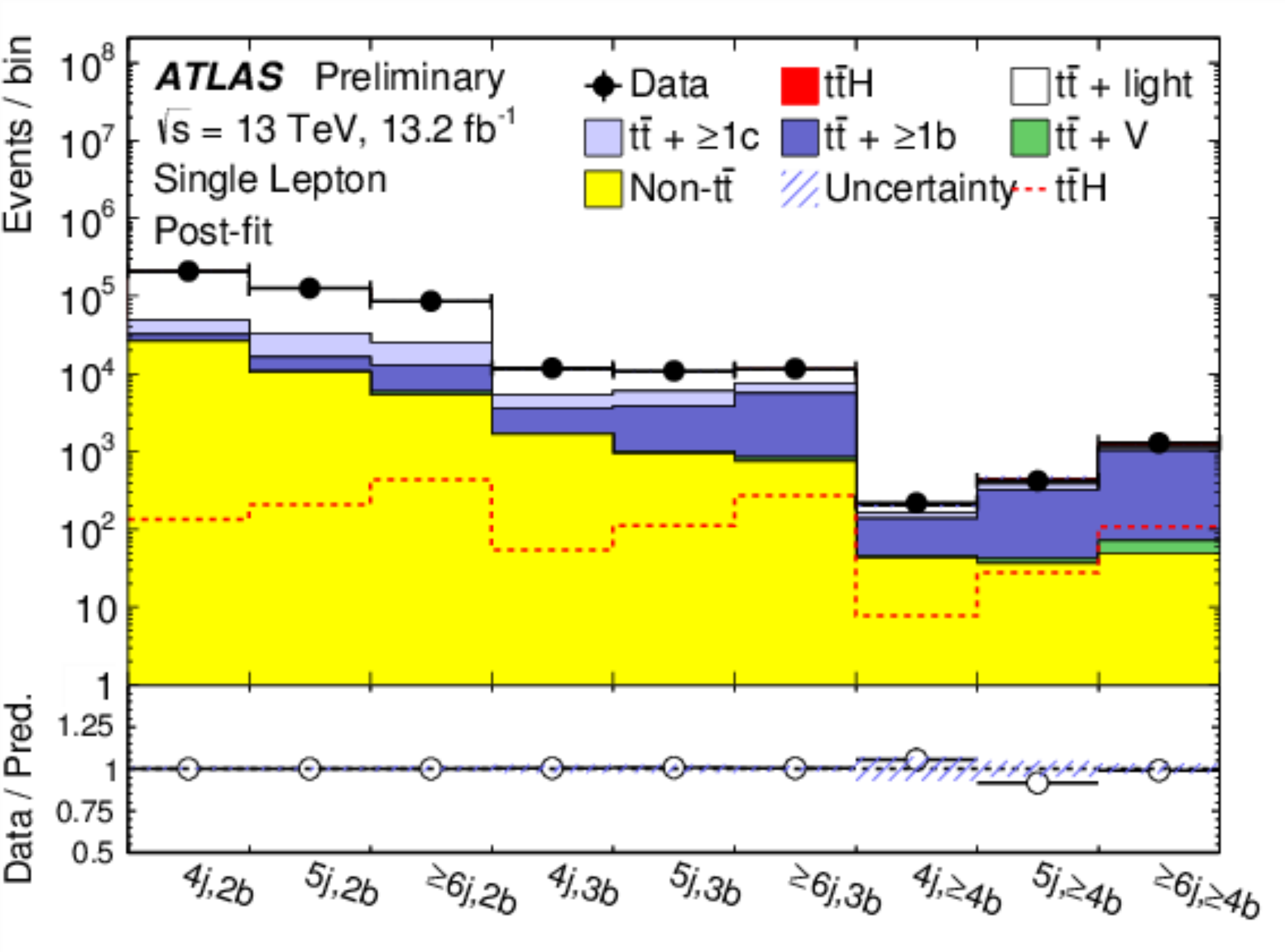}
  }
\caption[]{The distribution of events in jet categories in ATLAS
  before (left) and after (right) the fit.}
\label{fig:tth}
\end{figure}

Figure~\ref{fig:tth} shows the distribution of numbers of events in
bins of the number of jets and the number of b-tagged jets in ATLAS
before and after the fit. Significant mismodelling can be seen - much
larger that the signal size. CMS and ATLAS both used PowHeg v. 2 to
simulate \tt\bb, but CMS used a custom tune based on 8~TeV data and
get a better pre-fit agreement.

The fits are to BDT or Matrix Element Method (MEM) distributions. Uncertainties on background shapes are taken from scale variations,
PDFs etc., but both ATLAS and CMS add additional uncertainty to the
rate of top pairs with additional heavy-flavour quarks. The fit to the
data constrains the
modelling so that finally there is good agreement across the
distribution and much reduced uncertainties. This is essential if the
analysis is to be sensitive.

\begin{figure}
  \centerline{
    \includegraphics[height=0.43\linewidth]{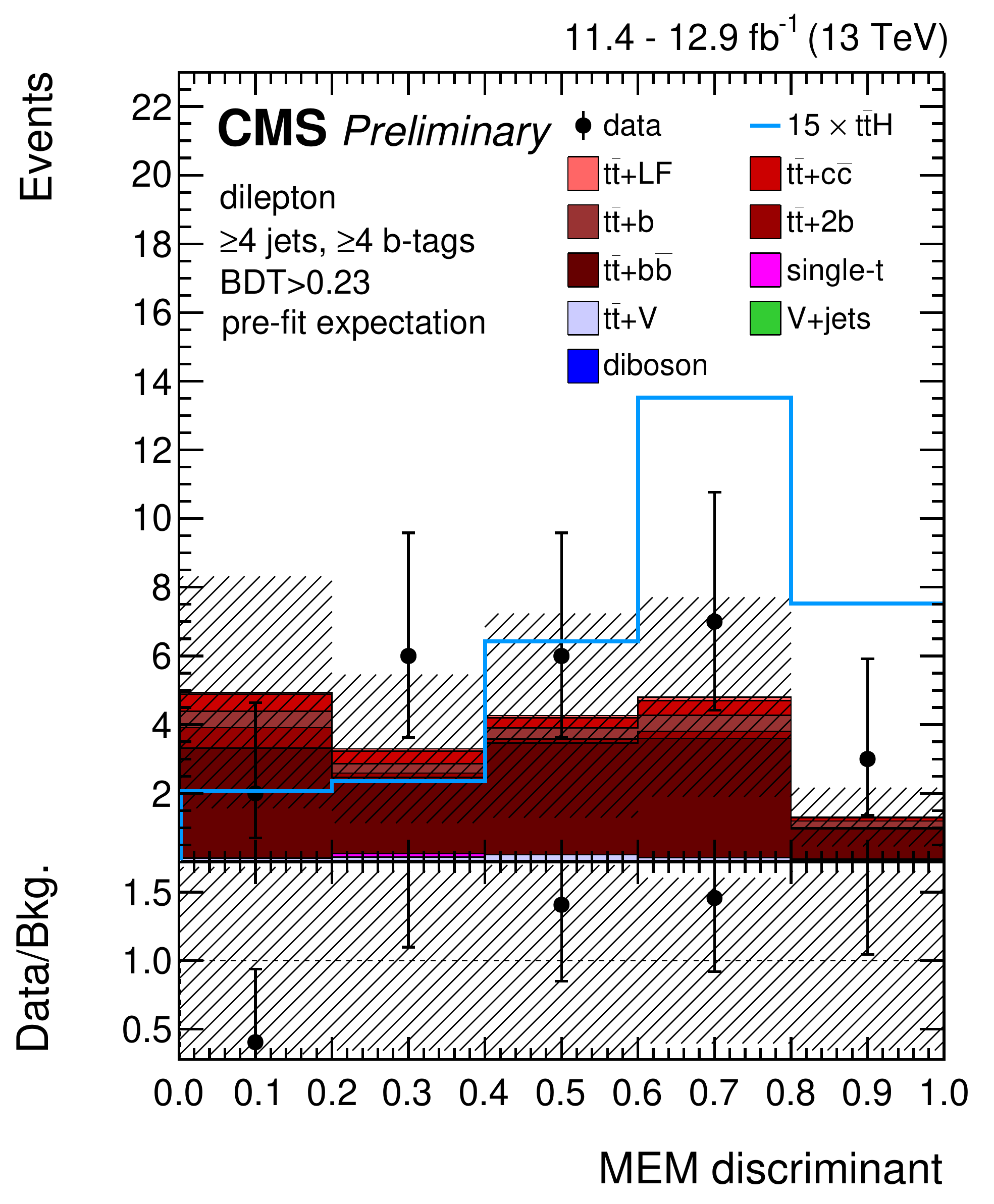},
    \includegraphics[height=0.43\linewidth]{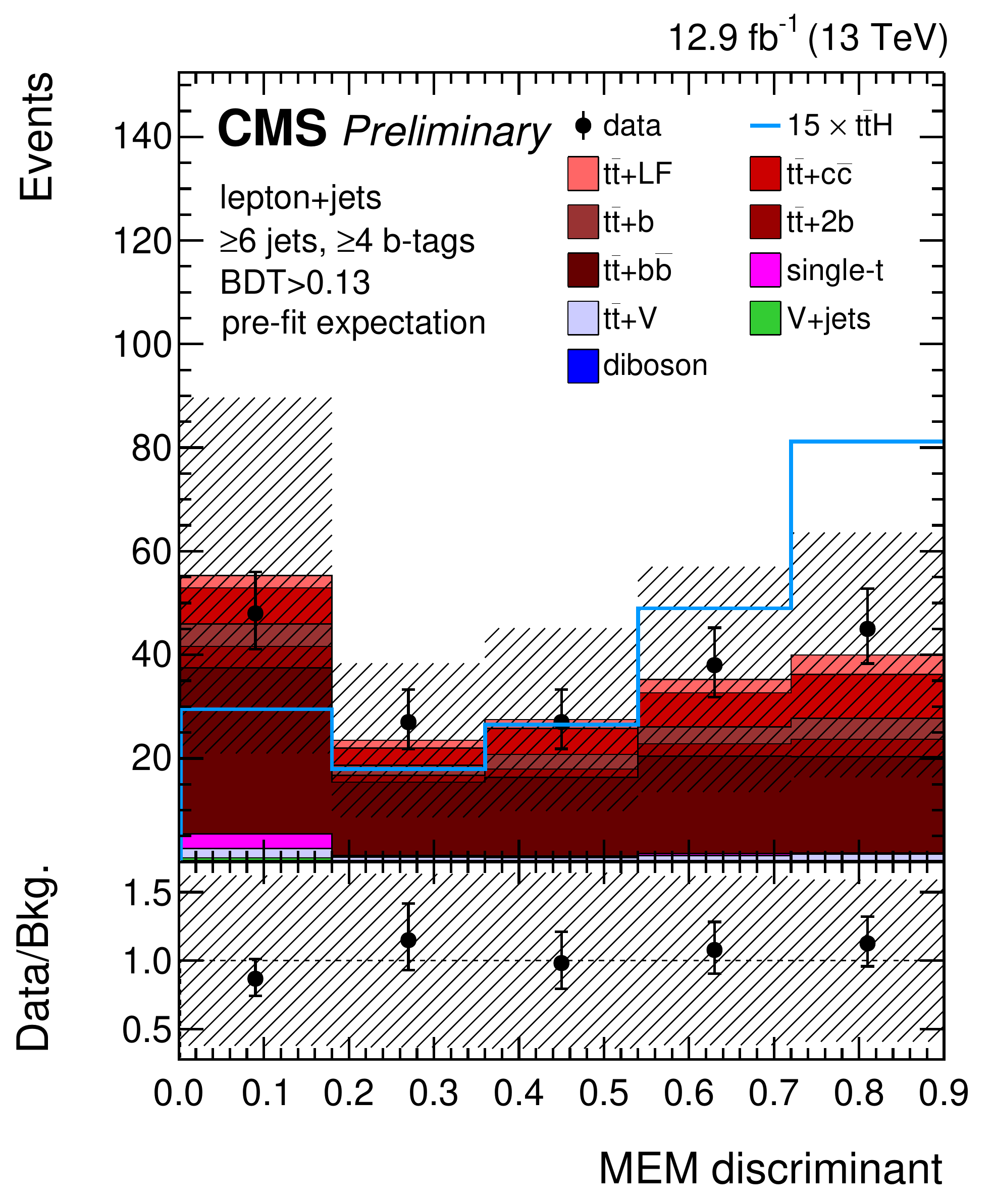},
    \includegraphics[height=0.43\linewidth]{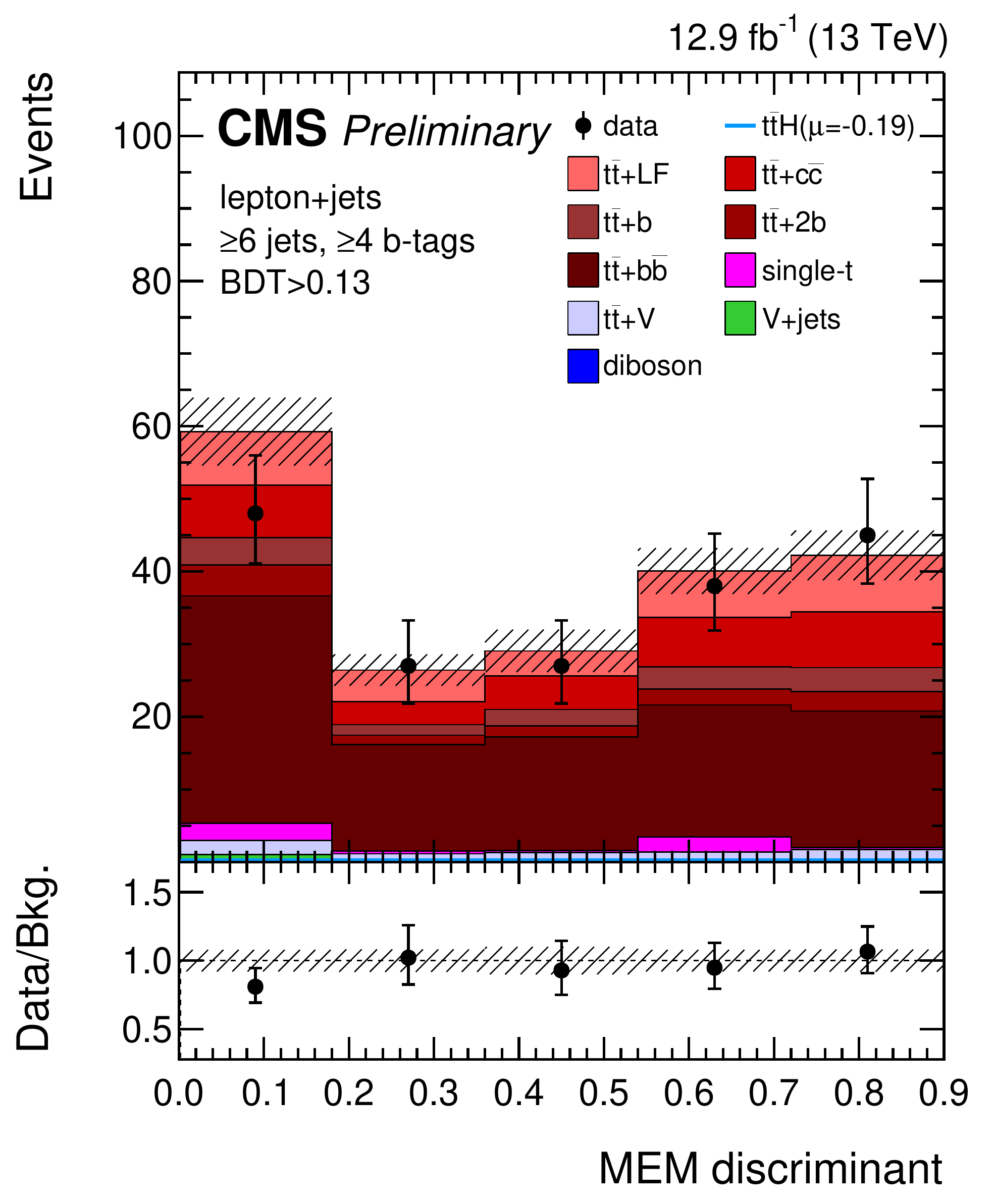}
  }
\caption[]{The most sensitive regions of the CMS \ttH\ search in di-lepton
  and single-lepton (centre) channels. The plots are shown pre-fit,
  and the single-lepton distribution is also shown post-fit (right).}
\label{fig:tth-final}
\end{figure}

Figure~\ref{fig:tth-final} shows the most signal-enriched regions of the CMS
analysis, in one and two lepton modes. The impact of the fit in
reducing the uncertainties can again be seen.

\section{Summary and Conclusions}
\label{sec:conclusions}

\begin{table}[ht]
\caption[]{The measurements of the \hbb\ rate from LHC run 2 in units of the
  standard model expectation.}
\label{ta:rates}
\vspace{0.4cm}
\begin{center}
\begin{tabular}{|c|c|r|c|}
\hline
 Production mode & Expt.  & Luminosity & Rate
\\ \hline
\ttH & ATLAS & 13.2~\ifb & 2.1  $^{+1.0}_{-0.9}$ \\
\ttH &   CMS & 12.9~\ifb & $-0.19\pm 0.80$\\
$VH$ & ATLAS & 13.2~\ifb & $0.21\pm 0.51$\\
VBF  &   CMS &  2.3~\ifb & $-3.7^{+2.4}_{-2.5}$\\
VBF+$\gamma$  &  ATLAS &  12.6~\ifb & $-3.9^{+2.8}_{-2.7}$\\
\hline
\multicolumn{2}{|c|}{Naive average} & n.a. & 0.2$\pm0.4$ \\
\hline
\end{tabular}
\end{center}
\end{table}

The measured rates of \hbb\ in the various analyses discussed here are
presented in table~\ref{ta:rates}, together with an average which
symmetrises the error bars and combines all the estimates as uncorrelated.
The run 2 data do not yet show clear evidence for the
\hbb\ process, but the precision is interesting. Future measurements will have much larger samples, which will not only
improve the statistical precision but also allow better control of the
systematics and in many cases the identification of higher purity regions.
The current, murky, situation will improve dramatically.


\section*{References}

\begin{thebibliography}{99}
\bibitem{ac2012}ATLAS and CMS collaborations,  \Journal{\JHEP}{8}{45}{2016}.

\bibitem{bdt} L. Breiman, J.H. Friedman, R.A. Olshen and C.J. Stone,
  {\em Classification and Regression Trees}, Wadsworth, Stamford, 1984.


\bibitem{cms-th}CMS Collaboration, CMS-PAS-HIG-16-019, 2016.

\bibitem{cms-vbf}CMS Collaboration, CMS-PAS-HIG-16-03, 2016.

\bibitem{atlas-vbf}ATLAS Collaboration, ATLAS-CONF-2016-063, 2016.

\bibitem{atlas-vh}ATLAS Collaboration, ATLAS-CONF-2016-091, 2016.

\bibitem{cms-tth}CMS collaboration, CMS-PAS-HIG-16-038, 2016.

\bibitem{atlas-tth}ATLAS collaboration,  ATLAS-CONF-2016-080, 2016.

\end{thebibliography}
\end{document}



